\documentstyle[12pt]{article}
\newcommand{\intx}{\int d^4 \! x \, }

\newcommand{\es}{\\[2mm]}

\newcommand{\journal}[4]{{\em #1~}#2\,(19#3)\,#4;}

\newcommand{\np}{\journal {Nucl. Phys.}}
\newcommand{\pl}{\journal {Phys. Lett.}}

\makeatletter
\@addtoreset{equation}{section}
\makeatother

\def\LP{\displaystyle{\Biggl(}}
 \def\wti{\widetilde}
\def\RP{\displaystyle{\Biggr)}}

\newcommand{\G}{\Gamma}
\newcommand{\D}{\Delta}
\renewcommand{\a}{\alpha}
\renewcommand{\b}{\beta}
\renewcommand{\d}{\delta}
\newcommand{\e}{\varepsilon}
\newcommand{\f}{\phi}
\newcommand{\fma}{\phi_{+}}
\newcommand{\fme}{\phi_{-}}

\newcommand{\g}{\gamma}

\newcommand{\x}{\xi}
\renewcommand{\l}{\lambda} \renewcommand{\L}{\Lambda}
\newcommand{\m}{\mu}
\newcommand{\n}{\nu}
\newcommand{\mn}{{\mu\nu}}

\newcommand{\p}{\psi}

\newcommand{\r}{\rho}
\newcommand{\s}{\sigma} \renewcommand{\S}{\Sigma}
\renewcommand{\t}{\theta}

\newcommand{\WT}{\wti T}
\newcommand{\Wl}{\wti \l}

\newcommand{\FF}{{\cal F}}

\newcommand{\RR}{{\cal R}}
\newcommand{\SS}{{\cal S}}
\newcommand{\TT}{{\cal T}}

\newcommand{\VV}{V}

\newcommand{\complex}{{\kern .1em {\raise .47ex
\hbox {$\scriptscriptstyle |$}}
    \kern -.4em {\rm C}}}
\newcommand{\real}{{{\rm I} \kern -.19em {\rm R}}}
\newcommand{\rational}{{\kern .1em {\raise .47ex
\hbox{$\scripscriptstyle |$}}
    \kern -.35em {\rm Q}}}
\renewcommand{\natural}{{\vrule height 1.6ex width
.05em depth 0ex \kern -.35em {\rm N}}}

\newcommand{\half}{\frac 1 2}
\newcommand{\pa}{\partial}

\newcommand{\sla}{\raise.15ex\hbox{$/$}\kern -.57em}

\newcommand{\twiddle}{\lower.9ex\rlap{$\kern -.1em\scriptstyle\sim$}}

\newcommand{\grad}{\nabla}

\newcommand{\eq}{\begin{equation}}
\newcommand{\eqn}[1]{\label{#1}\end{equation}}
\newcommand{\eea}{\end{eqnarray}}
\newcommand{\eqa}{\begin{eqnarray}}
\newcommand{\eqan}[1]{\label{#1}\end{eqnarray}}
\newcommand{\ba}{\begin{array}}
\newcommand{\ea}{\end{array}}
\newcommand{\eqac}{\begin{equation}\begin{array}{rcl}}
\newcommand{\eqacn}[1]{\end{array}\label{#1}\end{equation}}

\begin{document}


{\ }

\vspace{20mm}

\vspace{2cm}
\centerline{\LARGE Supersymmetric generalization of the}
\vspace{2mm}

\centerline{\LARGE tensor matter fields }  \vspace{2mm}

\vspace{9mm}

\centerline{V.E.R. Lemes
\footnote{e-mail address: dicktor@cbpfsu1.cat.cbpf.br},
A.L.M.A. Nogueira\footnote{e-mail address: nogue@cbpfsu1.cat.cbpf.br}}
\vspace{1mm}
\centerline{and}
\vspace{1mm}
\centerline{J.A.Helay\"{e}l-Neto
\footnote{e-mail address: helayel@cbpfsu1.cat.cbpf.br}}
\vspace{3mm}
\centerline{\it C.B.P.F}
\centerline{\it Centro Brasileiro de Pesquisas F\'{\i}sicas,}
\centerline{\it Rua Xavier Sigaud 150, 22290-180 Urca}
\centerline{\it Rio de Janeiro, Brazil}
\vspace{10mm}

\vspace{4mm}
\centerline{{\normalsize {\bf CBPF-NF 054/95}} }

\vspace{4mm}
\vspace{10mm}

\centerline{\Large{\bf Abstract}}\vspace{2mm}
\noindent
The supersymmetric generalization of a recently proposed Abelian axial
gauge model with antisymmetric tensor matter fields is presented.

\setcounter{page}{0}
\thispagestyle{empty}

\vfill
\pagebreak

\section{Introduction}

A few years ago, a line of investigation has been proposed by Avdeev 
and Chizhov \cite{ac} that consists in treating skew-symmetric rank-2 
tensor fields as matter rather than gauge degrees of freedom. The model 
studied in Ref.\cite{ac} has been further reassessed from the point of 
view of renormalization in the framework of BRS quantization \cite{nos}. 
In view of the potential relevance of matter-like tensor fields for 
phenomenology \cite{ac}, it is our purpose in this paper to discuss 
some facts concerning the formulation of an $N=1$ supersymmetric 
Abelian gauge model realizing the coupling of gauge fields to
matter tensor fields and their partners. One intends here to present a
superspace formulation of the model and exploit the possible relevance 
of extra bosonic supersymmetric partners (complex scalars) for the 
issue of symmetry breaking. We would like to mention that the 
supersymmetrization of 2-forms that appear as gauge fields is already 
known in connection with supergravity, and the so-called linear 
superfields appear to be the most appropriate multiplets to accomodate
the 2-form gauge fields \cite{Gates}. In our case, we aim at the 
supersymmetrization of matter 2-form fields coupled to Abelian gauge 
fields and their supersymmetric partners. 

 The present work is outlined as follows: in Section $2$, one searches 
for the supermultiplet that accomodates the matter tensor field and 
discusses its self-interaction; the coupling to the gauge supermultiplet 
is pursued in Sections $3$ and $4$; in Section $5$, one couples the 
well-known O' Raifeartaigh model \cite{rai} to the tensor-field 
supermultiplet and discusses some features concerning spontaneous 
symmetry and 
supersymmetry breaking. Finally, General Conclusions are drawn in 
Section $6$.

\section{Supersymmetrizing the tensor field}

Adopting conventions for the spinor algebra and the superspace parametrization 
of Ref.\cite{olie}, one finds that the superfield accomodating the 
skew-symmetric rank-$2$ tensor amongst its components is a spinor multiplet 
subject to the chirality constraint:

\eq\ba{rl}
\S_{a} =  & \p_{a} + \t^{b}\L_{ba} + \t^{2}\FF_{a} - i \t^{c}\s^{\m}_{c\dot{c}}
 \bar{\t}^{\dot{c}} \pa_{\m} \p_{a} \es
& - i \t^{c}\s^{\m}_{c\dot{c}} \bar{\t}^{\dot{c}} \pa_{\m} \t^{b} \L_{ba}
- \frac{1}{4} \t^{2}  \bar{\t}^{2} \pa_{\m}\pa^{\m} \p_{a}  ,
\ea\eqn{sig}
\eq\ba{rl}
\overline{\S}_{\dot{a}} = & \overline{\p}_{\dot{a}} + \bar{\t}_{\dot{b}}
\bar{\L}^{\dot{b}}_{\;\,\dot{a}} + \bar{\t}^{2} \overline{\FF}_{\dot{a}}
+ i \t^{c}\s^{\m}_{c\dot{c}}  \bar{\t}^{\dot{c}} \pa_{\m}
\overline{\p}_{\dot{a}} \es
& + i \t^{c}\s^{\m}_{c\dot{c}}  \bar{\t}^{\dot{c}} \pa_{\m} \bar{\t}_{\dot{b}}
\bar{\L}^{\dot{b}}_{\;\,\dot{a}} - \frac{1}{4} \t^{2}  \bar{\t}^{2}
\pa_{\m}\pa^{\m} \overline{\p}_{\dot{a}} ,
\ea\eqn{sigba}

\eq\ba{l}
\overline{D}_{\dot{b}}\S_{a} = D_{b}\overline{\S}_{\dot{a}} = 0,
\ea\eqn{chiralidade}

where $\p_{a}$ and $\FF_{a}$ are chiral spinors and $\L_{ba}$,
$\bar{\L}_{\dot{b}\dot{a}}$ are decomposed as:
\eq\ba{l}
     \L_{ba} = \e_{ba}\r + \s^{\mn}_{ba} \l_{\mn} \ ,  \es
     \bar{\L}_{\dot{b}\dot{a}} = - \e_{\dot{b}\dot{a}} \r^{\ast} -
\overline{\s}^{\mn}_{\dot{b}\dot{a}} \l^{\ast}_{\mn} \ .
\ea\eqn{deflambda}
According to the chiral properties of the superfield $\S_{a}$, the 
$\l_{\mn}$-tensor corresponds to the $(1,0)$-representation of Lorentz group. 
On the other hand, $\l^{\ast}_{\mn}$ yields the $(0,1)$-representation. We 
then write:
\eq\ba{l}
\l_{\mn} = T_{\mn} - i \WT_{\mn} \ , \es
\l^{\ast}_{\mn} = T_{\mn} + i \WT_{\mn} \ ,
\ea\eqn{osT}
where $\WT_{\mn} = \half \e_{\m\n\a\b} T^{\a\b}$. Notice also that $\Wl_{\mn} =
i \l_{\mn}$ and $\widetilde{(\l_{\mn}^{\ast})} = -i \l_{\mn}^{\ast}$, where the 
twiddle stands for the dual.

The canonical dimensions of the component fields read as below:
\eq\ba{l}
d(\p) = d(\overline{\p}) = \frac{1}{2} \ \es
d(\r) = d(\l_{\mn}) = 1 \ \es
d(\FF) =  d(\overline{\FF}) = \frac{3}{2} \ .
\ea\eqn{graus}
Based on dimensional arguments, we propose the following superspace action for
the $\S_{a}$-superfield:

\eq
\SS = \intx d^{2}\t  d^{2}\overline{\t} \hspace{.2cm} \frac{-1}{32}\{
D^{a}\S_{a} \overline{D}_{\dot{a}}\overline{\S}^{\dot{a}} +
q\S^{a}\S_{a}\overline{\S}_{\dot{a}}\overline{\S}^{\dot{a}} \}.
\eqn{superacao}

To check whether such an action is actually the supersymmetric extension of 
the model that treats $T_{\mn}$ as a matter field \cite{ac}, we have now to 
write down eq.($2.7$) in terms of the component fields 
$\p, \r, \l_{\mn}$ and $\FF$:

\eq\ba{rl}
\SS & = \intx \LP + \pa^{\m}\r \pa_{\m}\r^{\ast} - 16 \pa^{\m}\l_{\mn}
\pa_{\a}\l^{\ast\a\n} + i \overline{\FF}^{\dot{a}} \overline{\s}^{\m}_{\dot{a}
a} \pa_{\m} \FF^{a} - i \overline{\p}^{\dot{a}} \overline{\s}^{\m}_{\dot{a} a}
\pa_{\m} \pa^{\n}\pa_{\n} \p^{a} \es
& - q \frac{1}{\sqrt{2}}(\r^{2} - 4 \l_{\mn}\l^{\mn} )
\frac{1}{\sqrt{2}}(\r^{\ast 2} - 4 \l^{\ast}_{\a\b}\l^{\ast \a\b} ) +4q
\l^{\mn}\l_{\mn}\overline{\FF}_{\dot{a}}\overline{\p}^{\dot{a}} +4q
\l^{\ast\mn}\l_{\mn}^{\ast}\FF^{a}\p_{a} \es
& - 2q \FF^{a}\p_{a}\overline{\FF}_{\dot{a}}\overline{\p}^{\dot{a}} -
q\r^{2}\overline{\FF}_{\dot{a}}\overline{\p}^{\dot{a}} -
q(\r^{\ast})^{2}\FF^{a}\p_{a} + \frac{q}{2}\p^{a}\p_{a}\pa^{\m}\pa_{\m}
(\overline{\p}_{\dot{a}}\overline{\p}^{\dot{a}}) \es & - i
q\r\p^{a}\s^{\m}_{a\dot{a}}\pa_{\m}(\overline{\p}^{\dot{a}}\r^{\ast}) + 4 q
\r\p^{a}\s_{a\dot{a}}^{\m}\pa_{\b}(\l^{\ast\b}_{\:\;\:\;\m}
\overline{\p}^{\dot{a}})  - 4q \l_{\m\b}\pa^{\b}(\r^{\ast}
\overline{\p}_{\dot{a}})\overline{\s}^{\m \dot{a}a} \p_{a}  \es  & -16iq
\l_{\m\a}\pa_{\b}(\l^{\ast\b\a}\overline{\p}_{\dot{a}})
\overline{\s}^{\m\dot{a} a} \p_{a}  \RP.
\ea\eqn{acaocom}

Using $\l_{\mn} = \frac{1}{4} ( T_{\mn} - i \WT_{\mn} )$, we arrive at the 
relation:
\eq
16 \pa^{\a}\l_{\a\m}\pa_{\b}\l^{\ast\b\m} = 2 \pa^{\a}T_{\a\m}\pa_{\b}T^{\b\m}
- \half \pa^{\a}T^{\mn}\pa_{\a}T_{\mn} \; .
\eqn{bilinear}
The action above displays the terms proposed by Avdeev et al. in Ref.\cite{ac};
besides the anti-symmetric tensor, there appear a complex scalar and a pair of
spinors as its supersymmetric partners: $ \p_{a}$ , a non - physical fermion,
and $\FF_{a}$ , corresponding to a physical Weyl spinor. The undesirable 
presence of a spinor with lower canonical dimension ( $\frac{1}{2}$ , instead 
of $\frac{3}{2} $ ) generating, as expected, a higher derivative term in the 
Lagrangian can be avoided by a reshuffling of the spinorial degrees of 
freedom, if one keeps the interactions turned off. In fact, one can join both 
$\p_{a}$ and $\FF_{a}$ in a single fundamental Dirac spinor, $\Psi $, as 
follows:
\eq
\Psi \; = \; 
\left(
\ba{c}
\FF_{a}(x) \\
\overline\s^{\m\dot{a}a}\partial_{\m}\; \psi_{a}(x)
\ea
\right) \; .
\eqn{OneDirac}
The usual kinetic term, $ i\; \overline\Psi\;\gamma^{\m}\;\partial_{\m}\; 
(\Psi) $ , provides the kinetical terms for $\psi_{a}$ and $\FF_{a}$, turning 
the higher derivative term $, -i\;\overline\psi^{\dot{a}}\overline
\s^{\m}_{\dot{a}a}\partial_{\m}\partial^{2}
\psi^{a} $, into a matter of choice for the field basis. Nevertheless, this 
is true only for the free theory. The interaction sector of (\ref{acaocom}) 
cannot be re-expressed in terms of the Dirac spinor $\Psi $ , imposing a 
dissociation back to Weyl spinorial degrees of freedom. Therefore, it happens 
that the full theory must carry a higher derivative term, giving birth to a 
conjecture that this might be the fermionic counterpart of problems concerning 
the transverse sector of the original - bosonic -  model for the tensor matter 
field with interactions, as discussed in \cite{ac}. 

\section{The gauging of the model}
In order to perform the gauging of the model described by eq.(2.7), one 
proceeds along the usual lines and introduces a chiral scalar superfield, 
$\L$, to act as the gauge parameter:
\eq
\L = ( 1 - i \t^{a}\s^{\m}_{a\dot{a}}\overline{\t}^{\dot{a}}\pa_{\m} -
\frac{1}{4} \t^{2} \overline{\t}^{2} \pa_{\m}\pa^{\m} ) ( \f  + \t^{b}w_{b} +
\t^{2}\pi )
\eqn{invsemb}
\eq
\overline{\L} = ( 1 + i \t^{a}\s^{\m}_{a\dot{a}}\overline{\t}^{\dot{a}}\pa_{\m}
- \frac{1}{4} \t^{2} \overline{\t}^{2} \pa_{\m}\pa^{\m} ) ( \f^{\ast} +
\overline{\t}_{\dot{b}}\overline{w}^{\dot{b}} + \overline{\t}^{2}
\overline{\pi} ).
\eqn{invcomb}
The infinitesimal gauge transformations of the superfields $\S$ and
$\overline{\S}$ read as:
\eq\ba{rl}
\d \S_{a} &= ih \L \hspace{.1cm}\S_{a} \es
\d \overline\S_{\dot{a}} &= -ih \overline{\L}\hspace{.1cm}
\overline{\S}_{\dot{a}},
\ea\eqn{invsigma}
and the behaviour of $(D^{a}\S_{a})$ and
$(\overline{D_{\dot{a}}}\hspace{.1cm}\overline{\S^{\dot{a}}})$ under finite
transformations become:
\eq\ba{rl}
D^{a}\hspace{.1cm}\S_{a}^{'} &= e^{ih\L}\hspace{.1cm}(D^{a}\hspace{.1cm}\S_{a}
+ ih D^{a}\L\hspace{.1cm} \S_{a}) \es
\overline{D_{\dot{a}}}\hspace{.1cm}\overline{\S^{'\dot{a}}} &=
e^{-ih\overline{\L}}(\overline{D_{\dot{a}}}\hspace{.1cm}\overline{\S^{\dot{a}}}
-ih\overline{D_{\dot{a}}}\hspace{.1cm}\overline{\L}\hspace{.1cm}
\overline{\S^{\dot{a}}}).
\ea\eqn{noninv}
To gauge-covariantize the superspace derivatives, one introduces a gauge
connection superfield:
\eq
D_{a} \rightarrow \nabla_{a} = D_{a} + ih \G_{a},
\eqn{dericova}
in such a way that $\G_{a}$ transforms like
\eq
\G_{a}^{'} = \G_{a} - D_{a}\L .
\eqn{invgamma}
This yields:
\eq
(\nabla^{a}\S_{a})^{'} = e^{ih\L} (\nabla^{a}\S_{a}).
\eqn{superinv}
To achieve a U$(1)$-invariant action, one proposes
\eq
\SS = \intx d^{2}\t d^{2}\overline{\t} \LP
\grad^{a}\S_{a}\hspace{.1cm}e^{h\VV}\hspace{.1cm}\overline{\grad_{\dot{a}}}
\overline{\S^{\dot{a}}} \RP,
\eqn{gaugeaction}
where \VV \hspace{1mm} is the real scalar superfield \cite{jose} that
accomplishes the gauging of supersymmetric QED \cite{jose2}:
\eq
\VV^{'} = \VV + i\hspace{.1cm}( \overline{\L} - \L ).
\eqn{invV}
At this point, the gauge sector displays more degrees of freedom than 
what is actually required to perform the gauging. There are component 
vector fields in both $\G_{a}$ and $\VV$. However, we notice that the 
superfield $\G_{a}$ is not a true independent gauge potential. Indeed,
\eq
\G_{a} = - i D_{a} \VV
\eqn{relacomv}
reproduces correctly the gauge tranformation of $\G_{a}$ and, at the 
same time, eliminates the redundant degrees of freedom that would be 
otherwise present, if we were to keep $\G_{a}$ and $\VV$ as gauge 
superfields. Therefore, the locally U$(1)$- invariant action takes 
over the form:
\eq\ba{rl}
\SS & = \intx d^{2}\t \hspace{.2cm} \frac{-1}{128}  \LP
\overline{D}^{2}(e^{-\VV} D^{a} e^{\VV})\overline{D}^{2}(e^{-\VV} D_{a}
e^{\VV}) \RP \es & + \intx d^{2}\t  d^{2}\overline{\t}
\hspace{.2cm}\frac{-1}{32} \LP
\grad^{a}\S_{a}\hspace{.1cm}e^{h\VV}\hspace{.1cm}
\overline{\grad_{\dot{a}}}\overline{\S^{\dot{a}}} +
q  \S^{a}\S_{a}e^{2h\VV}\overline{\S}_{\dot{a}}\overline{\S}^{\dot{a}} \RP,
\ea\eqn{gaugeaction}
where
\eq\ba{rl}
\grad^{a}\S_{a} &= D^{a}\S_{a} + hD^{a}\VV \hspace{.1cm}\S_{a} \es
 \hspace{2mm} \overline{\grad_{\dot{a}}}\overline{\S^{\dot{a}}} &=
\overline{D_{\dot{a}}}\overline{\S^{\dot{a}}} + h\overline{D_{\dot{a}}}
\VV\hspace{.1cm}\overline{\S^{\dot{a}}}.
\ea\eqn{superderi}

The $\t$-expansion for the superfield $\VV$ brings about the following
component fields:
\eq\ba{rl}
\VV &= C + \t^{a}b_{a} + \overline{\t}_{\dot{a}}\overline{b}^{\dot{a}} + \t^{a}
\overline{\t}^{\dot{a}}\s^{\m}_{a\dot{a}}A_{\m} \es
& + \hspace{.1cm}\t^{2}\l + \overline{\t}^{2}\overline{\l} +
\t^{2}\overline{\t}_{\dot{a}}\overline{\g}^{\dot{a}} +
\overline{\t}^{2}\t^{a}\g_{a} + \t^{2}\overline{\t}^{2}\D,
\ea\eqn{expancao}
where C, $\l$, $\overline{\l}$ and $\D$ are scalars $b_{a}$ and $\g_{a}$ are
spinors and $A_{\m}$ is the U($1$)-gauge field. The gauge transformation of
these fields read as below:
\eq\ba{rl}
\d C &= i( \f^{\ast} - \f) ,\hspace{.2cm} \d\l = -i\pi,\hspace{.2cm}
\d\overline{\l} = i\overline{\pi},\hspace{.2cm} \es
\d b_{a} &= -iw_{a},\hspace{.2cm} \d \overline{b}_{\dot{a}} =
i\overline{w}_{\dot{a}} \es
\d\D &= \frac{i}{4}\pa^{\m}\pa_{\m} (\f - \f^{\ast}),\hspace{.2cm} \d A^{\m} =
-\pa^{\m}(\f + \f^{\ast}),\hspace{.2cm} \es
\d\g_{a} &=
\frac{1}{2}\s^{\m}_{a\dot{a}}\pa_{\m}\overline{w}^{\dot{a}},\hspace{.2cm}
\d\overline{\g}_{\dot{a}} = - \frac{1}{2}\overline{\s}_{\dot{a}a}\pa_{\m}w^{a}.
\ea\eqn{compinv}
As already known, for the sake of component-field calculations, one usually
works in the so-called Wess-Zumino gauge, where C, $b_{a}$ and $\l$ are gauged
away. The expansion for the exponential of the gauge superfield simplifies, 
in this gauge, according to:
\eq
e^{h\VV} = 1 + h\t^{a}\s^{\m}_{a\dot{a}}\overline{\t}^{\dot{a}} A_{\m} +
h\t^{2}\overline{\t}_{\dot{a}}\overline{\g}^{\dot{a}} +
h\overline{\t}^{2}\t^{a}\g_{a} + h\t^{2}\overline{\t}^{2}\D + \frac{1}{4}
h^{2}\t^{2}\overline{\t}^{2}A^{\m}A_{\m}.
\eqn{expe}
Using this gauge, the transformations of the matter fields are:
\eq\ba{rl}
\p_{a} &= ih\f\p_{a},\hspace{.2cm} \d\r = ih\f\r,\hspace{.2cm} \d\l_{\mn} =
ih\f\l_{\mn},\hspace{.2cm}
\d\FF_{a} = ih( \f\FF_{a} );
\ea\eqn{invmatter}
we get thereby the following transformations for the components $T_{\mn}$ and
$\WT_{\mn}$:
\eq
\d T_{\mn} = h\f \WT_{\mn}, \hspace{.5cm}\d \WT_{\mn} = - h\f T_{\mn}
\eqn{invqui}
These are precisely the Abelian gauge transformations for the tensor field as
firstly proposed in Ref.\cite{ac}.

\section{Component-field action in the Wess-Zumino gauge}
Having adopted the component fields as defined in the previous sections,
lengthy algebraic computations yield the following action in the Wess-Zumino
gauge:
\begin{displaymath}\ba{rl}
\SS &= \intx   \LP -\frac{1}{4}F^{\mn}F_{\mn} +2 \D^{2}  + i
\overline{\g}^{\dot{a}}\overline{\s}^{\m}_{\dot{a}a} \pa_{\m}\g^{a}
 + \pa^{\m}\r\pa_{\m}\r^{\ast} - 16 \pa^{\m}\l_{\mn}  \pa_{\a}\l^{\ast\a\n} \es
& + i \overline{\FF}^{\dot{a}} \overline{\s}^{\m}_{\dot{a} a} \pa_{\m} \FF^{a}
- i \overline{\p}^{\dot{a}} \overline{\s}^{\m}_{\dot{a} a} \pa_{\m}
\pa^{\n}\pa_{\n} \p^{a} - i\frac{h}{2}\pa^{\m}\r A_{\m}\r^{\ast} + i\frac{h}{2}
\r A^{\m}\pa_{\m}\r^{\ast} \es & + 2h\r\D\r^{\ast} + \frac{h^{2}}{4}\r
A^{\m}A_{\m}\r^{\ast} - h\g^{a}\FF_{a}\r^{\ast} -
h\r\overline{\g}_{\dot{a}}\overline{\FF}^{\dot{a}} -
i\frac{h}{2}\r\g^{a}\s^{\m}_{a\dot{a}}\pa_{\m}\overline{\p}^{\dot{a}} \es &
-i\frac{h}{2}
\r^{\ast}\overline{\g}^{\dot{a}}\overline{\s}^{\m}_{\dot{a}a}\pa_{\m}\p^{a} + i
\frac{h}{2} \pa_{\m}\r^{\ast}\p^{a}\s^{\m}_{a\dot{a}}\overline{\g}^{\dot{a}} -
i\frac{h}{2} \pa_{\m}\r\g^{a}\s^{\m}_{a\dot{a}}\overline{\p}^{\dot{a}}
- \frac{h}{4}
\p^{a}\s^{\m}_{a\dot{a}}A_{\m}\pa^{\n}\pa_{\n}\overline{\p}^{\dot{a}} \es & +
\frac{h}{4}
\overline{\p}^{\dot{a}}\overline{\s}^{\m}_{\dot{a}a}A_{\m}\pa^{\n}
\pa_{\n}\p^{a} + i h \D\overline{\p}^{\dot{a}}\overline{\s}^{\m}_{\dot{a}a}
\pa_{\m}\p^{a} + i h \D\p^{a}\s^{\m}_{a\dot{a}}\pa_{\m}\overline{\p}^{\dot{a}}
-\frac{h}{4}\pa^{\n}A_{\n}\overline{\p}^{\dot{a}}\overline{\s}^{\m}_{\dot{a}a}
\pa_{\m}\p^{a} \es & +
\frac{h}{4}\pa_{\n}A^{\n}\p^{a}\s^{\m}_{a\dot{a}}\pa_{\m}
\overline{\p}^{\dot{a}} + \frac{h}{2} \FF^{a}\s^{\m}_{a\dot{a}}A_{\m}
\overline{\FF}^{\dot{a}} - \frac{h}{8}\p^{a}(\s^{\n}\overline{\s}^{\m}
\s^{\a})_{a\dot{a}}
F_{\m\n}\pa_{\a}\overline{\p}^{\dot{a}} \es
 &+
\frac{h}{8}\pa_{\m}\p^{a}(\s^{\m}\overline{\s}^{\a}\s^{\n})_{a\dot{a}}
F_{\n\a}\overline{\p}^{\dot{a}}
 + h \r^{\ast}F_{\mn}\l^{\mn} + h\r F_{\m\n}\l^{\ast\m\n} \es
& + 2ih ( 4 \pa_{\m}\l^{\ast\m\n}\l_{\n\a}A^{\a} - 4
\pa^{\m}\l_{\mn}A_{\a}\l^{\ast\n\a}) \es
&- i
\frac{h^{2}}{2}\g^{a}(\s^{\m}\overline{\s}^{\n}\s^{\a})_{a\dot{a}}
\l_{\mn}A_{\a}\overline{\p}^{\dot{a}} + \frac{h}{2}\p^{a}(\s^{\n}
\overline{\s}^{\m}\s^{\a})_{a\dot{a}}
\overline{\g}^{\dot{a}}\pa_{\n} \l^{\ast}_{\m\a} -
\frac{h}{2}\pa_{\n}\p^{a}(\s^{\n}\overline{\s}^{\m}\s^{\a})_{a\dot{a}}
\l^{\ast}_{\m\a} \overline{\g}^{\dot{a}}\es
&+ i\frac{h^{2}}{2}
\overline{\g}^{\dot{a}}(\overline{\s}^{\m}\s^{\a}
\overline{\s}^{\b})_{\dot{a}a}A_{\m}\p^{a}\l^{\ast}_{\a\b} - \frac{h}{2}
\g^{a}(\s^{\a}\overline{\s}^{\b}\s^{\m})_{a\dot{a}}\pa_{\m}\l_{\a\b}
\overline{\p}^{\dot{a}} + \frac{h}{2}\g^{a}(\s^{a}
\overline{\s}^{\b}\s^{\m})_{a\dot{a}}\l_{\a\b}\pa_{\m}\overline{\p}^{\dot{a}}
\es & + \frac{h}{2}\pa_{\m}\p^{a}(\s^{\m}
\overline{\s}^{\n}\s^{\a})_{a\dot{a}}A_{\n}\pa_{\a}\overline{\p}^{\dot{a}}
+ \frac{h}{4}\pa_{\m}\p^{a}(\s^{\n}
\overline{\s}^{\m}\s^{\a})_{a\dot{a}}A_{\n}\pa_{\a}\overline{\p}^{\dot{a}} \es
&+ \frac{h}{4}
\pa_{\m}\p^{a}(\s^{\m}\overline{\s}^{\a}\s^{\n})_{a\dot{a}}A_{\n}\pa_{\a}
\overline{\p}^{\dot{a}} + h^{2}\D\p^{a}\s^{\m}_{a\dot{a}}A_{\m}
\overline{\p}^{\dot{a}} +\frac{i}{16}h^{2}\p^{a}(\s^{\n}\overline{\s}^{\m}
\s^{\a})_{a\dot{a}}
F_{\mn}A_{\a}\overline{\p}^{\dot{a}} \es
& + i
\frac{h^{2}}{16}\p^{a}(\s^{\n}\overline{\s}^{\m}\s^{\a})_{a\dot{a}}
F_{\a\m}A_{\n}\overline{\p}^{\dot{a}}  -i\frac{h^{2}}{2}\g^{a}(\s^{\a}
\overline{\s}^{\b}\s^{\m})_{a\dot{a}}
\l_{\a\b}A_{\m}\overline{\p}^{\dot{a}} \; + 
\ea\end{displaymath}
\eq\ba{rl}
& -
i\frac{h^{2}}{4}\pa_{\m}\p^{a}(\s^{\m}\overline{\s}^{\n}
\s^{\a})_{a\dot{a}}A_{\n}A_{\a}\overline{\p}^{\dot{a}} - i
\frac{h^{2}}{8}\pa_{\m}\p^{a}(\s^{\n}\overline{\s}^{\m}
\s^{\a})_{a\dot{a}}A_{\n}A_{\a}\overline{\p}^{\dot{a}} \es
& + i \frac{h^{2}}{4}
\p^{a}A_{\n}A_{\m}(\s^{\n}\overline{\s}^{\m}\s^{\a})_{a\dot{a}}
\pa_{\a}\overline{\p}^{\dot{a}} + i \frac{h^{2}}{8}\p^{a}A_{\n}A_{\m}
(\s^{\n}\overline{\s}^{\a}
\s^{\m})_{a\dot{a}}\pa_{\a}\overline{\p}^{\dot{a}} \es & +
\frac{h^{3}}{8}\p^{a}(\s^{\n}\overline{\s}^{\a}\s^{\m})_{a\dot{a}}
A_{\n}A_{\a}A_{\m}\overline{\p}^{\dot{a}} - h^{2}\g^{a}\p_{a}
\overline{\g}_{\dot{a}}\overline{\p}^{\dot{a}} + 4 h^{2}A^{\n}A_{\m}
\l_{\n\b}\l^{\ast\b\m} \es
&- q \frac{1}{\sqrt{2}}(\r^{2} - 4 \l_{\mn}\l^{\mn} )
\frac{1}{\sqrt{2}}(\r^{\ast 2} - 4 \l^{\ast}_{\a\b}\l^{\ast \a\b} ) +4q
\l^{\mn}\l_{\mn}\overline{\FF}_{\dot{a}}\overline{\p}^{\dot{a}} +4q
\l^{\ast\mn}\l_{\mn}^{\ast}\FF^{a}\p_{a} \es
& - 2q \FF^{a}\p_{a}\overline{\FF}_{\dot{a}}\overline{\p}^{\dot{a}} -
q\r^{2}\overline{\FF}_{\dot{a}}\overline{\p}^{\dot{a}} -
q(\r^{\ast})^{2}\FF^{a}\p_{a} + \frac{q}{2}\p^{a}\p_{a}\pa^{\m}\pa_{\m}
(\overline{\p}_{\dot{a}}\overline{\p}^{\dot{a}}) \es &-
iq\r\p^{a}\s^{\m}_{a\dot{a}}\pa_{\m}(\overline{\p}^{\dot{a}}\r^{\ast}) + 4q
\r\p^{a}\s^{\m}_{a\dot{a}}\pa^{\b}(\l^{\ast}_{\b\m}\overline{\p}^{\dot{a}}) -
4q
\l_{\m\b}\pa^{\b}(\r^{\ast}\overline{\p}_{\dot{a}})
\overline{\s}^{\m\dot{a}a}\p_{a} \es &- 16i q \l_{\m\a}
\pa_{\b}(\l^{\ast\b\a}\overline{\p}_{\dot{a}})
\overline{\s}^{\m\dot{a}a}\p_{a} - q h \D\p^{a}\p_{a}
\overline{\p}_{\dot{a}}\overline{\p}^{\dot{a}} -
iq\frac{h}{2} A_{\m}\p^{a}\p_{a}\pa^{\m}(\overline{\p}_{\dot{a}}
\overline{\p}^{\dot{a}}) \es
& + iq\frac{h}{2}
A_{\m}\pa^{\m}(\p^{a}\p_{a})\overline{\p}_{\dot{a}}\overline{\p}^{\dot{a}}
- q \frac{h^{2}}{2}
A^{\m}A_{\m}\p^{a}\p_{a}\overline{\p}_{\dot{a}}\overline{\p}^{\dot{a}}
- q h \r\g^{a}\p_{a}\overline{\p}_{\dot{a}}\overline{\p}^{\dot{a}} - q h
\r^{\ast}\overline{\g}_{\dot{a}}\overline{\p}^{\dot{a}}\p^{a}\p_{a} \es
&- q
h\g^{a}\s^{\mn}_{ab}\p^{b}\overline{\p}_{\dot{a}}
\overline{\p}^{\dot{a}}\l_{\mn} + q h \overline{\g^{\dot{a}}}
\overline{\s}^{\mn}_{\dot{a}\dot{b}}\l^{\ast}_{\mn}
\overline{\p}^{\dot{b}}\p^{a}\p_{a} - q h \r\p^{a}
\s^{\m}_{a\dot{a}}A_{\m}\overline{\p}^{\dot{a}}\r^{\ast} \es
& - 2i q h
\r\p^{a}A_{\m}(\s^{\m}\overline{\s}^{\a}
\s^{\b})_{a\dot{a}}\l^{\ast}_{\a\b}\overline{\p}^{\dot{a}} + 2 i q h
\r^{\ast}\p^{a}(\s^{\b}\overline{\s}^{\a}\s^{\m})_{a\dot{a}}
A_{\m}\l_{\a\b}\overline{\p}^{\dot{a}} \es & + 2 q h \p^{a}(\s^{\a}
\overline{\s}^{\b}\s^{\g}
\overline{\s}^{\m}\s^{\n})_{a\dot{a}}\l_{\a\b}A_{\g}\l^{\ast}_{\m\n}
\overline{\p}^{\dot{a}} \RP.    \ea\eqn{wsacao}
We should stress here a remarkable difference with respect to the case 
of the chiral and anti-chiral \underline{scalar} superfields (Wess-Zumino 
model \cite{WZM}), namely, the minimal coupling of $\S$ and $\overline{\S}$ 
to the gauge sector necessarily affects the $\S$ - superfield 
self-interaction terms as one reads off from eq ($3.11$). The gauging of 
the $U(1)$ - symmetry enriches the self-interactions of the tensor field 
not only through its fermionic supersymmetric partners, but also through 
the introduction of the gauge boson and the gaugino at the level of the matter 
self-interaction terms (these new couplings are compatible with 
power-counting renormalizability). This is so because the
model presented here is based on a single spinor superfield. Had we 
introduced a couple of spinor superfields, $\S_{a}$ and $\TT_{a}$, 
with opposite $U(1)$
charges:
\eq\ba{rl}
\S^{'}_{a} &= e^{ih\L}\S_{a}, \es
\TT^{'}_{a} &= e^{-ih\L}\TT_{a},
\ea\eqn{taus}
a self-interacting term of the form
$(\S^{a}\TT_{a}\overline{\S}_{\dot{a}}\overline{\TT}^{\dot{a}})$ would
automatically be invariant whenever the symmetry is gauged, and there would be
no need for introducing the vector superfield to ensure local invariance. Such
a mixed self-interacting term could, in principle, be thought of as a possible
source for a mass term for the spinor superfields, whenever the physical scalar
component $\r$ develops a non-trivial vacuum expectation value. Nevertheless,
by analysing the $\r$ - field interactions in the scalar potential, one
concludes that there is no room for spontaneous  symmetry breaking as induced
by such a component field (and, similarly, for its counterpart inside $\TT$ ). 
On the other hand, we could think to introduce a gauge-invariant mass term of 
the form
\eq
\SS_{mass} = \intx (d^{2}\t \hspace{.2cm}i\frac{m}{16}\S^{a}\TT_{a} -
d^{2}\overline{\t} \hspace{.2cm}i\frac{m}{16}
\overline{\S}_{\dot{a}}\overline{\TT}^{\dot{a}});
\eqn{qusemass}
however, a mixed mass term like the one above introduces two massive
excitations of the type $k^{4} = m^{4}$ in the
spectrum. So, regardless the sign of $m^{2}$, a tachyon shall always be
present; hence such a mass term is disregarded.
\section{On Spontaneous Symmetry Breaking}
Due to the spinorial character of the superfield $\S_{a}$, it cannot be used 
to accomplish a spontaneous breaking. Indeed, Lorentz invariance is
lost whenever $\S_{a}$ acquires a non-trivial vacuum expectation value. The
idea in the present section is to couple, in a gauge-invariant manner, the
well-known O' Raifeartaigh model \cite{rai} to the spinor superfield
$\S_{a}$, so as to understand the issue of an eventual mass generation for 
$\S_{a}$ via spontaneous internal symmetry or supersymmetry breakingdown. 
For the sake of concreteness, the model we adopt to discuss such a
matter is defined by the action below:
\eq\ba{rl}
\SS &= \intx d^{2}\t d^{2}\overline{\t} \LP \overline{\f}\f +
\overline{\fma}e^{h\VV}\fma + \overline{\fme}e^{-h\VV}\fme \RP \es
&+ \intx d^{2}\t \LP \half m\f^{2} + \m \fma\fme + f \f + g \f\fma\fme
+ G\S^{a}\S_{a}\fme\fme  \RP \es
&+ \intx  d^{2}\overline{\t} \LP \half m\overline{\f}^{2} + \m \overline{\fma}
\overline{\fme} + f \overline{\f} + g \overline{\f} \overline{\fma}
\overline{\fme} + G\overline{\S}_{\dot{a}}
\overline{\S}^{\dot{a}}\overline{\fme}
\overline{\fme} \RP \es
&+ \intx d^{2}\t \hspace{.2cm} \frac{-1}{128}  \LP \overline{D}^{2}(e^{-\VV}
D^{a} e^{\VV})\overline{D}^{2}(e^{-\VV} D_{a} e^{\VV}) \RP \es & + \intx
d^{2}\t  d^{2}\overline{\t} \hspace{.2cm}\frac{-1}{32} \LP
\grad^{a}\S_{a}\hspace{.1cm}e^{h\VV}\hspace{.1cm}\overline{\grad_{\dot{a}}}
\overline{\S^{\dot{a}}} + q  \S^{a}\S_{a}e^{2h\VV}\overline{\S}_{\dot{a}}
\overline{\S}^{\dot{a}} \RP,
\ea\eqn{oraios}
where the chiral scalar superfields $\f$, $\f_{+}$ and $\f_{-}$ are
parametrized as follows:
\eq\ba{rl}
\f &=  ( 1 - i \t^{a}\s^{\m}_{a\dot{a}}\overline{\t}^{\dot{a}}\pa_{\m} -
\frac{1}{4} \t^{2} \overline{\t}^{2} \pa_{\m}\pa^{\m} ) ( A + \t^{b}\x_{b} +
\t^{2}b ) \es
\fma &=  ( 1 - i \t^{a}\s^{\m}_{a\dot{a}}\overline{\t}^{\dot{a}}\pa_{\m} -
\frac{1}{4} \t^{2} \overline{\t}^{2} \pa_{\m}\pa^{\m} ) ( A_{+} + \t^{b}\x_{+b}
+ \t^{2}b_{+} ) \es
\fme &=  ( 1 - i \t^{a}\s^{\m}_{a\dot{a}}\overline{\t}^{\dot{a}}\pa_{\m} -
\frac{1}{4}\t^{2} \overline{\t}^{2} \pa_{\m}\pa^{\m} ) ( A_{-} + \t^{b}\x_{-b}
+ \t^{2}b_{-} ) \; ;
\ea\eqn{fis}
m and $\m$ are mass parameters, f has dimension of ${mass}^{2}$, whereas g and
G are dimensionless coupling constants.
$\S_{a}$ and $\f_{-}$ have opposite U$(1)$ - charges. This action, in terms of
components, reads:
\begin{displaymath}\ba{rl}
\SS &= \intx 4\LP 4\{ \pa^{\m}A^{\ast}\pa_{\m}A +
\pa^{\m}A^{\ast}_{-}\pa_{\m}A_{-} + \pa^{\m}A^{\ast}_{+}\pa_{\m}A_{+} \} + 4\{
b^{\ast}b + b^{\ast}_{-}b_{-} + b^{\ast}_{+}b_{+} \} \es
&+ 2i \{ \overline{\x}^{\dot{a}} \overline{\s}^{\m}_{\dot{a}a} \pa_{\m} \x^{a}
+ \overline{\x}^{\dot{a}}_{-} \overline{\s}^{\m}_{\dot{a}a} \pa_{\m} \x^{a}_{-}
+ \overline{\x}^{\dot{a}}_{+} \overline{\s}^{\m}_{\dot{a}a} \pa_{\m} \x^{a}_{+}
\} \RP \; +
\ea\end{displaymath}
\eq\ba{rl}
&+ \intx \LP 16h A^{\ast}_{+}\D A_{+} + 8ih A^{\ast}_{+}\pa^{\m}A_{\m}A_{+} +
4h^{2} A^{\ast}_{+}A^{\m}A_{\m}A_{+} - 8h A^{\ast}_{+}\g^{a}\x_{+a} \es
&- 8h\overline{\x}_{+\dot{a}}\overline{\g}^{\dot{a}}A_{+} + 4h
\x^{+a}\s^{\m}_{a\dot{a}}A_{\m}\overline{\x}^{\dot{a}}_{+} +
16ih\pa^{\m}A^{\ast}_{+}A_{\m}A_{+} \RP \es
&+ \intx \LP -16h A^{\ast}_{-}\D A_{-} - 8ih A^{\ast}_{-}\pa^{\m}A_{\m}A_{-} +
4h^{2}A^{\ast}_{-}A^{\m}A_{\m}A_{-} + 8h A^{\ast}_{-}\g^{a}\x_{-a} \es
&+ 8h\overline{\x}_{-\dot{a}}\overline{\g}^{\dot{a}}A_{-} - 4h
\x^{-a}\s^{\m}_{a\dot{a}}A_{\m}\overline{\x}^{\dot{a}}_{-} -
16ih\pa^{\m}A^{\ast}_{-}A_{\m}A_{-} \RP \es
&+ \intx \LP m(-4bA + \x^{a}\x_{a}) + m(-4b^{\ast}A^{\ast} +
\overline{\x}_{\dot{a}}\overline{\x}^{\dot{a}}) - 4fb - 4fb^{\ast} \RP \es
& + \intx 2\LP \m(-2b_{+}A_{-} - 2b_{-}A_{+} + \x^{a}_{+}\x_{a -}) +
\m(-2b^{\ast}_{+}A^{\ast}_{-} - 2b^{\ast}_{-}A^{\ast}_{+} +
\overline{\x}_{\dot{a}+}\overline{\x}^{\dot{a}}_{-}) \RP  \es
 & + \intx 2\LP g(-2bA_{+}A_{-} - 2Ab_{+}A_{-} - 2AA_{+}b_{-} + A\x^{a}_{+}\x_{a
-} + A_{-}\x^{a}\x_{a +} + A_{+}\x^{a}\x_{a -}) \es 
& + g (-2b^{\ast}A^{\ast}_{+}A^{\ast}_{-} - 2A^{\ast}b^{\ast}_{+}A^{\ast}_{-} -
2A^{\ast}A^{\ast}_{+}b^{\ast}_{-} + A^{\ast}\overline{\x}_{\dot{a}
+}\overline{\x}^{\dot{a} -} +
A^{\ast}_{-}\overline{\x}_{\dot{a}}\overline{\x}^{\dot{a}}_{+} +
A^{\ast}_{+}\overline{\x}_{\dot{a}}\overline{\x}^{\dot{a}}_{-}) \RP \es
& + \intx 4G\LP (-2\FF^{a}\p_{a} + \r^{2} +4 \l_{\mn}\l^{\mn})(A_{-})^{2} +
(-2\overline{\FF}_{\dot{a}}\overline{\p}^{\dot{a}} + (\r^{\ast})^{2} +4
\l^{\ast}_{\mn}\l^{\ast \mn})(A^{\ast}_{-})^{2} \es
& + 2\p^{b}\p_{b}(-4b_{-}A_{-} + \x^{a}_{-}\x_{a -}) +
2\overline{\p}_{\dot{b}}\overline{\p}^{\dot{b}}(-4b^{\ast}_{-}A^{\ast}_{-} +
\overline{\x}_{\dot{a} -}\overline{\x}^{\dot{a} -}) \es
& - 2(\r\p^{b} - \s^{\mn\,ba}\l_{\mn}\p_{a})(\x_{b -}A_{-})
+2(\r^{\ast}\overline{\p}_{\dot{b}} -
\l^{\ast}_{\mn}\overline{\s}^{\mn}_{\dot{b}\dot{a}}
\overline{\p}^{\dot{a}})(\overline{\x}^{\dot{b}}_{-} A^{\ast}_{-}) \RP \es
 & +  \intx   \LP  + 2h\r\D\r^{\ast}  +2 \D^{2}  + \frac{h^{2}}{4}\r
A^{\m}A_{\m}\r^{\ast} + \ldots  \RP ,
\ea \eqn{bAxi2}
where the dots stand for spinorial partners and derivative terms that are 
completely irrelevant for discussing spontaneous symmetry breaking. Also, 
the $\r $-field does not acquire a non-trivial v.e.v., as already mentioned 
in the previous section. The scalars that could, in principle, trigger 
spontaneous symmetry breaking are only $ A $, $ A_{+} $ and $ A_{-} $, if 
one starts from action (\ref{bAxi2}).  

The only possible way to endow the tensor field $ \l_{\m\n} $ with a mass, 
via internal symmetry or supersymmetry breaking, would be by means of a 
coupling of the form
\eq
\S^{a}\S_{a} \phi^{2} \; ,
\eqn{acopla}
as dictated by supersymmetry, through the chirality constraints on $ \S $ 
and $ \phi $. No matter the number of scalar superfields present in the 
model, whenever the breaking takes place and a mass is generated for 
$ \l_{\mu\nu} $ as a byproduct, one notices that both the imaginary part of 
$ \r $ and one longitudinal degree of freedom of $ \l_{\mu\nu} $ provide the 
spectrum with a tachyonic excitation, without any chance of avoiding this 
fact at the expenses of the $\Delta $-field coupling, enriched by an additional 
Fayet-Iliopoulos term \cite{Fayet} (actually, a $\Delta $-type term does not 
break supersymmetry whenever it is addded to the O' Raifeartaigh model. 
Moreover, a $\Delta $-term never couples to $ \l_{\mu\nu} $, since 
$ \l_{\mu\nu} \l^{\ast}_{\mu\nu} $ is identically vanishing). Therefore, our 
final conclusion is that the masslessness of $ \r $ and $ \l_{\mu\nu} $ 
cannot be avoided in a consistent way, just by invoking internal symmetry or 
supersymmetry breaking as realised by a set of scalar superfields. 

\section{General Conclusions}
The supersymmetrization of the matter tensor field first investigated in
Ref.{\cite{ac}} has been worked out here in terms of a spinor chiral
superfield, $\S_{a}$, whose kinetic and self-interacting terms have been found
in $N=1$ - superspace. The gauging of the model reveals some peculiarities,
such as the need of gauge fields to appear in the matter self-interactions.

Scalar degrees of freedom that accompany the fermionic partners of 
$ \l_{\mu\nu} $ cannot be the source for spontaneous symmetry or 
supersymmetry breaking, as it could in principle be thought. The reason is 
that Lorentz invariance prevents $\S_{a}$ from developing a non-trivial 
vacuum expectation value.

A thorough analysis of the coupling between $ \S_{a} $ and chiral and 
anti-chiral scalar superfields indicate that no spontaneous breaking takes 
place. In components, the scalar $ \r $ and the tensor $ \l_{\mu\nu} $ 
display a quartic coupling to the physical scalar components of the 
additional matter superfields, namely, $ \r^{2} A_{-}^{2} $ and 
$ \l_{\mu\nu}^{2} A_{-}^{2} $, and no non-trivial minimum with 
$<A_{-}> \;\neq 0 $ shows up (a non-trivial minimum with non-zero vacuum 
expectation value restricted to the neutral scalar A is possible, but has no 
consequence for mass generation). So, spontaneous breaking does not happen 
to be a possibility for inducing a mass for the tensor $ \l_{\mu\nu} $.

As a next step, the analysis of matter tensor fields in the framework of 
$ N=2 $ extended supersymmetries and the non-Abelian version of the $ N=1 $ 
model are to be pursued.  

\section{Conventions}
\eq
\s^{\m} = ( 1,\s ), \hspace{.5cm} \overline{\s}^{\m} = ( 1,-\s ), 
\hspace{.5cm} \overline{\s}^{\m}_{\dot{b}a} = \s^{\m}_{a\dot{b}}\;\; ,
\eqn{pauli}
where $\s = (\s_{1},\s_{2},\s_{3})$ are the Pauli matrices

\eq
\s_{1} = \left( \begin{array}{cc} 0 & 1 \\ 1 & 0 \end{array} \right) , 
\hspace{.5cm} 
\s_{2} = \left( \begin{array}{cc} 0 & -i \\ i & 0 \end{array} \right), 
\hspace{.5cm} 
\s_{3} = \left( \begin{array}{cc} 1 & 0 \\ 0 & -1 \end{array} \right) .
\eqn{dirac-pauli}
In addition, the matrices $\s^{\mn}$ and $\overline{\s}^{\mn}$ 
( $ (\frac{1}{2},0)$ and $ (0,\frac{1}{2})$ SO(1,3) generators) are given by
\eq\ba{rl}
\s^{\m}_{a\dot{a}}\overline{\s}^{\n \dot{a}b} &= \eta^{\mn}\d_{a}^{\; b} - i
(\s^{\mn})_{a}^{\; b} \es
\overline{\s}^{\m\dot{a} a}\s^{\n}_{a\dot{b}} &= \eta^{\mn}\d^{\dot{a}}_{\;
\dot{b}} - i (\overline{\s}^{\mn})^{\dot{a}}_{\; \dot{b}} \; ,
\ea\eqn{matrizes}
and the trace is
\eq\ba{rl}
\s^{\m}_{a\dot{a}}\overline{\s}^{\n\dot{a}b}\s^{\a}_{b\dot{b}}
\overline{\s}^{\b\dot{b}a} = 2\LP \eta^{\mn}\eta^{\a\b} -
\eta^{\m\a}\eta^{\n\b} + \eta^{\m\b}\eta^{\n\a} + i \e^{\m\n\a\b} \RP,
\ea\eqn{traco}
where $\e^{0123} = -\e_{0123} = 1$. \es
The summation convention is:
\eq
\t\eta = \t^{a}\eta_{a}, \hspace{.5cm} \overline{\t}\overline{\eta} =
\overline{\t}_{\dot{a}}\overline{\eta}^{\dot{a}}\; ,
\eqn{sobedesce}
where lowering and raising of indices are effected through
\eq
\t^{a} = \e^{ab}\t_{b}, \hspace{.5cm} \t_{a} = \e_{ab}\t^{b},
\eqn{lorai}
with $ \e_{ab} = - \e_{ba} $ , (the same for dotted indices).
Differentation with respect to the anticommuting parameters $\t_{a}$ ,
$\overline{\t}_{\dot{a}}$ is defined by
\eq
\frac{\pa}{\pa\t^{a}}\t^{b} = \d_{a}^{\, b} \hspace{.5cm}
\frac{\pa}{\pa\overline{\t}^{\dot{a}}}\overline{\t}^{\dot{b}} =
\d_{\dot{a}}^{\, \dot{b}}.
\eqn{delta1}
Covariant derivatives with respect to the supersymmetry transformations are:
\eq\ba{rl}
D_{a} & = \frac{\pa}{\pa\t^{a}} -
i\s^{\m}_{a\dot{a}}\overline{\t}^{\dot{a}}\pa_{\m} \es
\overline{D}_{\dot{a}} & = - \frac{\pa}{\pa\overline{\t}^{\dot{a}}} + i
\t^{a}\s^{\m}_{a\dot{a}}\pa_{\m},
\ea\eqn{superderivadas}
and they obey the anticommutation relations
\eq
\{ D_{a},\overline{D}_{\dot{a}} \} = 2 i \s^{\m}_{a\dot{a}}\pa_{\m} \; ;
 \hspace{.5cm} \{D_{a},D_{b} \} = 0 = \{ \overline{D}_{a},\overline{D}_{\dot{b}}
\}.
\eqn{relations}
The Dirac matrices, $ \gamma^{\m} $ , playing a role in the purely 
physical spinorial kinetic term, 
$ i\;\overline\Psi \gamma^{\m}\partial_{\m} \Psi $ , have the 
following expression in terms of the Pauli matrices:
\eq
\gamma^{\m} \; = \; 
\left( 
\ba{cc}
0 &  \s^{\m} \\
\overline\s^{\m} & 0
\ea
\right) \; .
\eqn{Gamma} 
The spinor $\overline\Psi $ is defined as usually:
\eq
\overline\Psi = \Psi^{\dagger}\gamma^{0}\; , \;\mbox{where} \;\;\gamma^{0} 
\; = \;
\left(
\ba{cc}
0 & 1 \\
1 & 0
\ea
\right) \; .
\eqn{Barrado}

\vspace{2cm}

\noindent{\large{\bf Acknowledgements}}

We wish to thank Dr. M. A. de Andrade for helpful discussions on an earlier
manuscript.
The {\it Conselho Nacional de Desenvolvimento Cientifico e Tecnologico},
$CNPq$-Brasil is gratefully acknowledged for the financial support.


\end{document}